\documentclass[11pt]{article}
\usepackage{dsfont}
\usepackage{cite}
\usepackage{hyperref}
\usepackage{enumitem,mathtools}
\usepackage{adjustbox}
\usepackage{authblk}
\usepackage{pdfrender}
\usepackage[outline]{contour}
\usepackage{tikz}
\usepackage{graphicx}
\usepackage{xifthen}
\usepackage{xargs}
\usepackage{amssymb}
\usepackage{mathtools}
\usepackage{wasysym}
\usepackage{accents}

\usetikzlibrary{decorations.pathreplacing, patterns}
\usetikzlibrary{calc}
\usetikzlibrary{arrows}
\usetikzlibrary{patterns}

\newtheorem{conjecture}{Conjecture}

\newtheorem{definition}{Definition}

\newcommand{\RR}{\mathds{R}}

\newcommand{\Tau}{\mathcal{T}}
\DeclareRobustCommand{\Uv}[1]{%
  \accentset{\mbox{\large\bfseries .}}{U}_{#1}}
\newcommand{\Ue}[1]{\bar{U}_{#1}}

\definecolor{bl}{RGB}{0,186,252}
\definecolor{gr}{RGB}{41,163,0}
\definecolor{re}{RGB}{238,48,0}
\definecolor{vi}{RGB}{126,48,255}
\usepackage{wrapfig}

\newlist{thmlist}{enumerate}{1}
\setlist[thmlist]{label=(\alph*), ref=\thethm.(\alph*),noitemsep}

\usepackage{thmtools}

\declaretheorem[
    name=Theorem,
    Refname={Theorem,Theorems}]{thm}

\usepackage{nameref,hyperref}
\usepackage[capitalize]{cleveref}

\Crefname{thm}{Theorem}{Theorems}
\Crefname{thmlisti}{Theorem}{Theorems}

\usepackage{pgf, tikz}
\usepackage{graphicx}
\usepackage{tikzsymbols}
\usepackage{vwcol}
\usepackage{fancyvrb}
\usepackage{float}
\usepackage[skip=6pt]{caption}
\usepackage{pgffor}
\usepackage{graphics}
\usepackage{stackengine}
\usepackage{multicol}

\definecolor{mapgrey}{rgb}{.6,.6,.6}
\definecolor{mapred}{rgb}{.95,.2,.2}
\definecolor{maporange}{rgb}{.9,.6,0}
\definecolor{mapgreen}{rgb}{0,.55,0}
\definecolor{mapdarkgreen}{rgb}{0,.4,.1}
\definecolor{mapblue}{rgb}{.45,.55,.8}
\definecolor{mapviolet}{rgb}{.6,0,.6}

\definecolor{lightred}{RGB}{255,0,130}
\definecolor{darkgreen}{RGB}{0,168,112}
\definecolor{pygreen}{RGB}{0,128,0}
\definecolor{lightorange}{RGB}{255,192,0}
\definecolor{darkred}{RGB}{210,0,43}
\definecolor{myorange}{RGB}{230,153,43}
\definecolor{mygreen}{RGB}{0,112,99}
\definecolor{myprovedgreen}{RGB}{0,178,26}
\definecolor{mycounterred}{RGB}{204,0,0}
\definecolor{mygrey}{RGB}{153,153,153}
\definecolor{myblue}{RGB}{76,102,204}
\definecolor{myviolet}{RGB}{153,0,153}

\title{\bf Density of triangulated ternary disc packings}
\author[1,2]{Thomas Fernique}
\author[2]{Daria Pchelina}
\affil[1]{CNRS}
\affil[2]{LIPN, Univ. Paris Nord, 93430 Villetaneuse, France}

\begin{document}
\maketitle
\begin{abstract}
  We consider \emph{ternary} disc packings of the plane, i.e. the
  packings using discs of three different radii. Packings in which
  each ``hole'' is bounded by three pairwise tangent discs are called
  \emph{triangulated}. There are 164 pairs $(r,s)$, $1{>}r{>}s$,
  allowing triangulated packings by discs of radii 1, $r$ and $s$.  In
  this paper, we enhance existing methods of dealing with
  maximal-density packings in order to find ternary triangulated
  packings which maximize the density among all the packings with the
  same disc radii.  We showed for 16 pairs that the density is
  maximized by a triangulated ternary packing; for 15 other pairs, we
  proved the density to be maximized by a triangulated packing using
  only two sizes of discs; for 40 pairs, we found non-triangulated
  packings strictly denser than any triangulated one; finally, we
  classified the remaining cases where our methods are not
  applicable. 
\end{abstract} 

\section{Density of disc and sphere packings}

Given a finite set $S$ of discs, a \emph{packing} of the plane by $S$ is a
collection of translated copies of discs from $S$ with disjoint
interiors.

Given a packing $P$, its \emph{density} $\delta(P)$ is the
proportion of the plane covered by the discs. More formally,
$$ \delta(P) := \limsup_{n\rightarrow\infty}\frac{\text{area}([-n,n]^2\cap P)}{\text{area}([-n,n]^2)} .$$

Nowadays, the density of disc packings is widely studied in different
contexts.  The worst-case optimal density of packings in triangular
and circular containers is found in~\cite{fTriangles,fDiscs}. In
computer science, there are various connections between sphere
packings and error-correcting codes~\cite{Conway1998}. Researchers in
chemical physics used Monte Carlo simulations on 2-disc packings and,
among others, obtained lower bounds on the maximal density of packings
with particular disc sizes~\cite{chem2d}. Two other groups of
physicists found lower bounds on maximal densities of packings in
$\RR^3$ with 2 sizes of spheres~\cite{phys3d,chem3d}. Upper bounds on
the density are usually much harder to obtain.

The main problem we are interested in is the following: given a finite
set of ball sizes in $\RR^2$ (or $\RR^3$), find a packing of the plane
(or of the space) maximizing the density.

\begin{figure}[htp]
  \begin{center}
    \begin{minipage}[p]{.67\linewidth}
      \includegraphics[width=\linewidth]{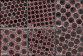}
    \end{minipage}\vskip1ex
    
    \begin{minipage}[p]{.67\hsize}
      \includegraphics[width=\linewidth]{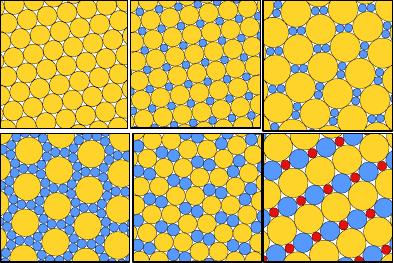}
    \end{minipage}
    \caption{Disc packings self-assembled from colloidal nanodiscs and
      nanorods in~\cite{chembinter} (at the top) which very accurately
      correspond to triangulated packings (on the bottom).}
    \label{fig:chemists}
  \end{center}
\end{figure}

Answering this question has a few practical applications. Chemists,
for example, are interested in the disc and sphere sizes maximizing
the density in order to eventually design compact materials using
spherical nanoparticles of given
sizes~\cite{chembinter,chem2d,phys3d}.  Figure~\ref{fig:chemists}
gives an illustration of experimental results from
\cite{chembinter}.

The first known studies of the densest packings go back to Kepler. Many
advances in this area have been made since then.

\subsection{Spheres}
\begin{figure}[h]
  \centering
  \begin{minipage}[p]{.43\linewidth}
    \includegraphics[width=\linewidth]{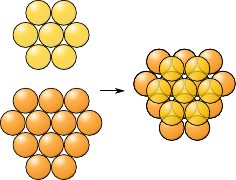}
    \caption{First step of construction of a 3D close-packing.}
    \label{fig:hex_3d}
  \end{minipage}\hfill
  \begin{minipage}[p]{.43\linewidth}
    \includegraphics[width=\linewidth]{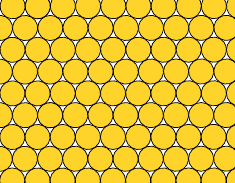}
    \caption{2D hexagonal packing.}
    \label{fig:hex}
  \end{minipage}
\end{figure}

In a Kepler manuscript dated by 1611, we find a description of the
``cannonball'' packing followed by an assertion that it is a densest
\emph{1-sphere packings} (i.e. packings by equally sized spheres) of
the three-dimensional Euclidean space. This assertion is widely known
by name of the Kepler conjecture.  The ``cannonball'' packing, also
called face-centered-cubic (FCC) packing, belongs to a family of
packings formed by stacking layers 
of spheres centered in the vertices of a triangular lattice, like it is
shown in Figure~\ref{fig:hex_3d}. After placing the first two layers,
at each step, there are two choices of how to place the next
layer. This gives us an uncountable set of packings having the same
density. These packings are called \emph{close-packings} of equal
spheres.

\begin{conjecture}[Kepler 1611]
  The density $\delta(P)$ of packing $P$ of $\RR^3$ by unit spheres
  never exceeds the density of a close-packing:
  \begin{equation}
    \delta(P)\leq\frac{\pi}{3\sqrt{2}}
    \,.
    \label{eq:kepler}
  \end{equation}
\end{conjecture}

The first advancement in a proof of the Kepler conjecture was made by
Gauss who, in 1831, showed that close sphere packings maximize the
density among all possible \emph{lattice} packings, i.e. those where
the disc centers form a lattice~\cite{gauss}. However, the proof of
the whole conjecture took four centuries to be found. Hilbert included
this conjecture, also named ``the sphere packing problem'', in his
famous list of 23 problems published in 1900.

The Kepler conjecture was finally proved in a series of 6 papers
submitted by Hales and Furgeson in 1998~\cite{halesDCG,hales}.  Their
computer-assisted proof took 8 years to be fully reviewed. In 2003,
Hales founded a project called Flyspeck in order to fully verify his
proof by an automated theorem prover. Flyspeck was completed in 2014
including the proof of the Kepler conjecture in the list of computer
verified proofs~\cite{halesformal}.

The rough idea of the proof consists of locally redistributing (or
weighting) the density function and showing
inequality~(\ref{eq:kepler}) for this redistributed (weighted)
density. Lagarias calls this approach
``localization''~\cite{lagarias02}. All in all, in our work, we use
the same general ideas which are discussed in detail in
Section~\ref{sec:strat}.

\subsection{Discs}
\subsubsection{1-disc packings}

The two-dimensional variant of the Kepler conjecture claims the 2D
hexagonal packing on the plane (see Fig.~\ref{fig:hex}) to have the
highest density among all planar packings by identical discs.

In 1772, Lagrange proved it to be a densest among lattice packings. The
general result was first shown by Thue in 1910~\cite{thue1910}.  His
proof was however considered incomplete, a reliable proof was given by
Fejes-T\'{o}th in 1942~\cite{toth}.

A packing by a set of discs is called \emph{saturated} if no more
discs from this set can be added to the packing without intersecting
already placed discs. In our setup, we always assume packings to be
saturated since we are interested in the upper bounds on the density
and adding discs to a packing augments it.

The proof of the two-dimensional Kepler conjecture contains the basics
of the strategy used to prove similar results for packings with
several disc sizes, like binary packings (discussed in the next
section) and ternary packings which are studied in this paper. We thus
find it useful to provide the idea of this proof, following its
version given in~\cite{simplethue}.

Let $P^*$ denote the hexagonal packing of identical discs of radius
1. Our aim is to show for any saturated two-dimensional packing $P$
using discs if radius 1, that its density does not exceed the density
$\delta^*:=\frac{\pi}{2\sqrt{3}}$ of $P^*$.

\begin{figure}[h]
  \centering
  \begin{minipage}[h]{.45\linewidth}
    \centering
    \includegraphics[width=.9\linewidth]{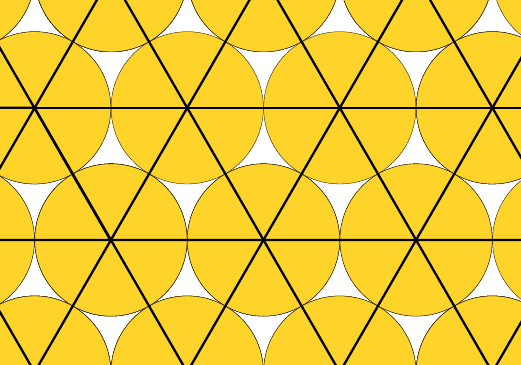}
    \caption{The Delaunay triangulation of the hexagonal packing $P^*$.}
    \label{fig:triang_hex}
  \end{minipage}\hfill
  \begin{minipage}[h]{.45\linewidth}
    \centering
    \includegraphics[width=.9\linewidth]{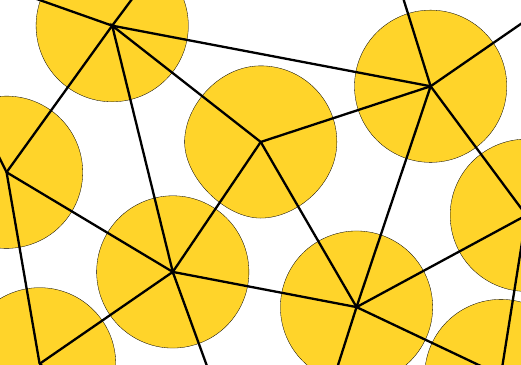}
    \caption{An example of a packing $P$ with its Delaunay triangulation.}
    \label{fig:triang_rand}
  \end{minipage}
\end{figure}

First, let us consider the Delaunay
triangulations\footnote{See~\cite{rourke}, especially chapter 23, for
  the definition and properties of Delaunay triangulations.} of disc
centers of $P^*$ and $P$ (see
Fig.~\ref{fig:triang_hex},~\ref{fig:triang_rand}).  Notice that in the
triangulation of $P^*$, all triangles are equilateral triangles of
side $2$. We define the density of a packing in a given triangle to be
equal to the proportion of this triangle covered by discs of the
packing. The density in any triangle of the triangulation of $P^*$
equals $\frac{\pi}{2\sqrt{3}}=\delta^*$.

It turns out that the density of any triangle in the triangulation of
$P$ is less or equal to $\delta^*$, as proven
in~\cite{simplethue}. This allows us to conclude.

This proof is rather simple due to its ``locality'': instead of
showing that the density of the whole packing $P$ is bounded by
$\delta^*$, we show it for each triangle of its triangulation (which
is a stronger assertion). Intuitively, the smaller are the units we
work on, the more ``local'' the proof is.

\vspace{1ex}

Packings of the plane where, as in the hexagonal one, each ``hole'' is
bounded by three pairwise tangent discs are called
\emph{triangulated}. More formally,
\begin{definition}
  A packing is called \textbf{triangulated} if the graph formed by
  connecting the centers of every pair of tangent discs is
  a triangulation.
\end{definition}

Fejes T\'{o}th in~\cite{tothcompact} called such packings ``compact'':
since saturated triangulated packings have no ``huge holes'', they
intuitively look the most compact. Moreover, around each disc, its
neighbors form a corona of tangent discs which looks like a locally
``optimal'' way to pack. For these reasons, triangulated packings
appear to be the best candidates to maximize the density on the whole
plane.

Notice, that for a fixed $n$, there exists only a finite number of
$n$-tuples of disc radii $(r_1,\cdots,r_n)$
s.t. $1{=}r_1{>}\cdots{>}r_n{>}0$ allowing a triangulated packing
where all $n$ disc sizes are present~\cite{messerschmidt21}.

\subsubsection{2-disc packings}

Let us now consider binary packings of the plane. We study the
following question: given two discs of radii $1$ and $r<1$, what is
the maximal density of a packing by copies of these discs? We can
always obtain $\frac{\pi}{2\sqrt{3}}$, the density of the hexagonal
packing, by using only one of these discs which gives as a lower bound
on the maximal density. Florian in~\cite{boundFlorian} derived an
upper bound on the density which is equal to the density in the
triangle formed by 2 small and one big pairwise tangent
discs. \cite{ferniqueBounds} gives tighter lower and upper bounds of
maximal density of binary packings of the plane, for all values of
$r\in(0,1)$.

\begin{figure}[h]\centering
  \includegraphics[width=.7\linewidth]{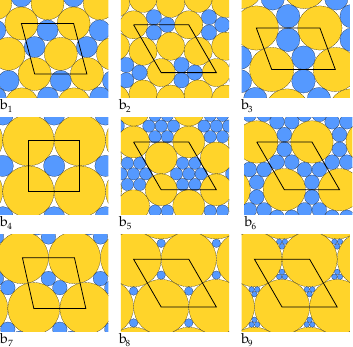}
  \caption{9 triangulated periodic binary packings maximizing the
    density among packings with the respective disc sizes.}
  \label{fig:9_triangulated}
\end{figure}\vskip1ex

There are 9 values of $r$ allowing triangulated binary packings where
the both disc sizes are
present~\cite{kennedy06}.
Such packings are shown in Fig.~\ref{fig:9_triangulated}. Each of the
depicted packings is \emph{periodic}, i.e. if $P$ is a packing in
question, there are two non co-linear vectors $u$ and $v$, called
periods, such that $P+u=P+v=P$. Notice that in this paper, we always
consider packings of the whole plane, and since the triangulated
packings we show here and below are all periodic, it is enough to
represent their fundamental domain (a parallelogram formed by the
period vectors, marked in black in Fig.~\ref{fig:9_triangulated}) to
see how the whole plane is packed.

Notice that for each of these values of $r$, there is actually an
infinite number of packings having the same density as the one
depicted in Fig.~\ref{fig:9_triangulated}.  First, changing a finite
portion of a packings does not affect its density. Moreover, for
$b_1$, $b_3$, and $b_7$, there exist non-periodic triangulated
packings with a different global structure, having the same density as
the ones from Fig.~\ref{fig:9_triangulated}~\cite{kennedy06}. For the sake of
simplicity, we choose to depict the periodic ones.

It turns out that for these 9 radii $r$, the density is maximized by a
triangulated binary packing -- namely, the ones shown in
Figure~\ref{fig:9_triangulated}~\cite{heppes00,heppes03,kennedy04,BF}.

This result suggests the following conjecture~\cite{ConConj}.

\begin{conjecture}[Connelly, 2018]
  If a finite set of discs allows a triangulated saturated packing,
  then the density of packings by these discs is maximized on a
  triangulated packing.
  \label{conj:connelly}
\end{conjecture}

This holds for 1-disc packings and 2-disc (\emph{binary})
packings. To study this conjecture, the next step is to verify it for
3-disc (\emph{ternary}) packings which was the main motivation of
our work.

\section{Result, plan of the paper}

Let us turn to the ternary packings. To begin with, we need to find
the sizes of discs allowing triangulated ternary packings. This
problem was solved in~\cite{164}: there are 164 pairs $(r,s)$
featuring triangulated packings with discs of radii $1,r,s$. In this
paper, the triplet of discs with radii associated to each of such
pairs is called a \emph{case}.

The ternary cases are indexed by positive integers from $1$ to $164$,
like in~\cite{164}.  To avoid confusion, the binary cases (pairs of
disc radii allowing binary triangulated packings) are denoted by
$b_1,\dots,b_9$ which respectively correspond to the cases 1--9
in~\cite{BF}.

The Connelly conjecture (Conjecture~\ref{conj:connelly}) is applicable
only to the cases having triangulated \emph{saturated} packings. This
eliminates 15 cases where no triangulated packing is saturated and
leaves us with 149 cases. 

Our main contribution is a classification of 71 cases formulated in
the following theorem:\\
    
\begin{thm}\label{thm:1}
\leavevmode
\makeatletter
\@nobreaktrue
\begin{thmlist}
\item  \label{thm:16} For the 16 following cases: 53, 54, 55, 56, 66, 76, 77, 79, 93, 108,
  115, 116, 118, 129, 131, 146, the density is maximized by a
  triangulated ternary packing.
\item \label{thm:15} For the cases 1--15, the density is maximized by
  triangulated binary packings. For cases 1--5, it is the triangulated
  packing of $b_8$; for case 6 --- $b_4$; for cases 7--9 --- $b_7$; for
  cases 10--16 --- $b_9$.
\item \label{thm:40} For the 40 following cases: 19, 20, 25, 47, 51, 60, 63, 64, 70,
  73, 80, 92, 95, 97, 98, 99, 100, 104, 110, 111, 117, 119, 126, 132,
  133, 135, 136, 137, 138, 139, 141, 142, 151, 152, 154, 159, 161,
  162, 163, 164, there exists a non-triangulated packing denser than
  any triangulated one.
  \end{thmlist}
  \label{th:16}
\end{thm}\vskip-2ex

\begin{figure}[htp]\centering
  \includegraphics[width=\linewidth]{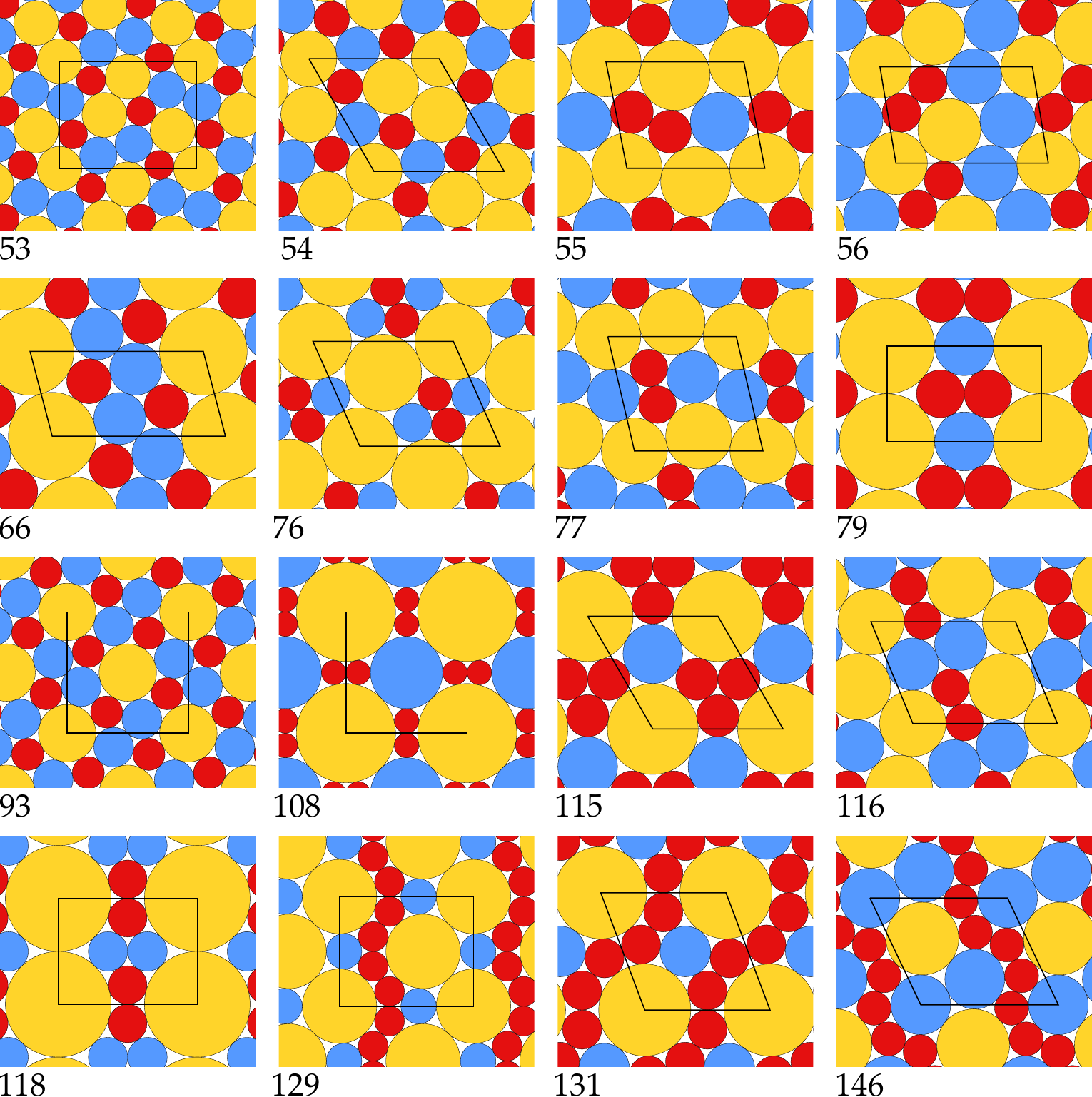}
  \caption{The 16 triangulated ternary packings proved to maximize the
    density (the numbers correspond to the numbering in~\cite{164}).}
  \label{fig:proved}
\end{figure}

The values of radii corresponding to the cases from
Theorem~\ref{thm:1} are given in~\cite{164}. The triangulated packings
maximizing the density for the cases from Th.~\ref{thm:16} are
depicted in Fig.~\ref{fig:proved}. For Th.~\ref{thm:15}, the binary
triangulated packings which maximize the density are present in
Fig.~\ref{fig:9_triangulated} while the ternary triangulated packings
are given in Fig.~\ref{fig:1-15}. Triangulated ternary packings and
non-triangulated binary denser packings for Th.~\ref{thm:40} are given
in Fig.~\ref{fig:73} and in the Appendix~\ref{app:counter}.

\begin{figure}[htp]\centering
  \includegraphics[trim=7pt 8pt 8pt 8pt,
  clip=true,width=\linewidth]{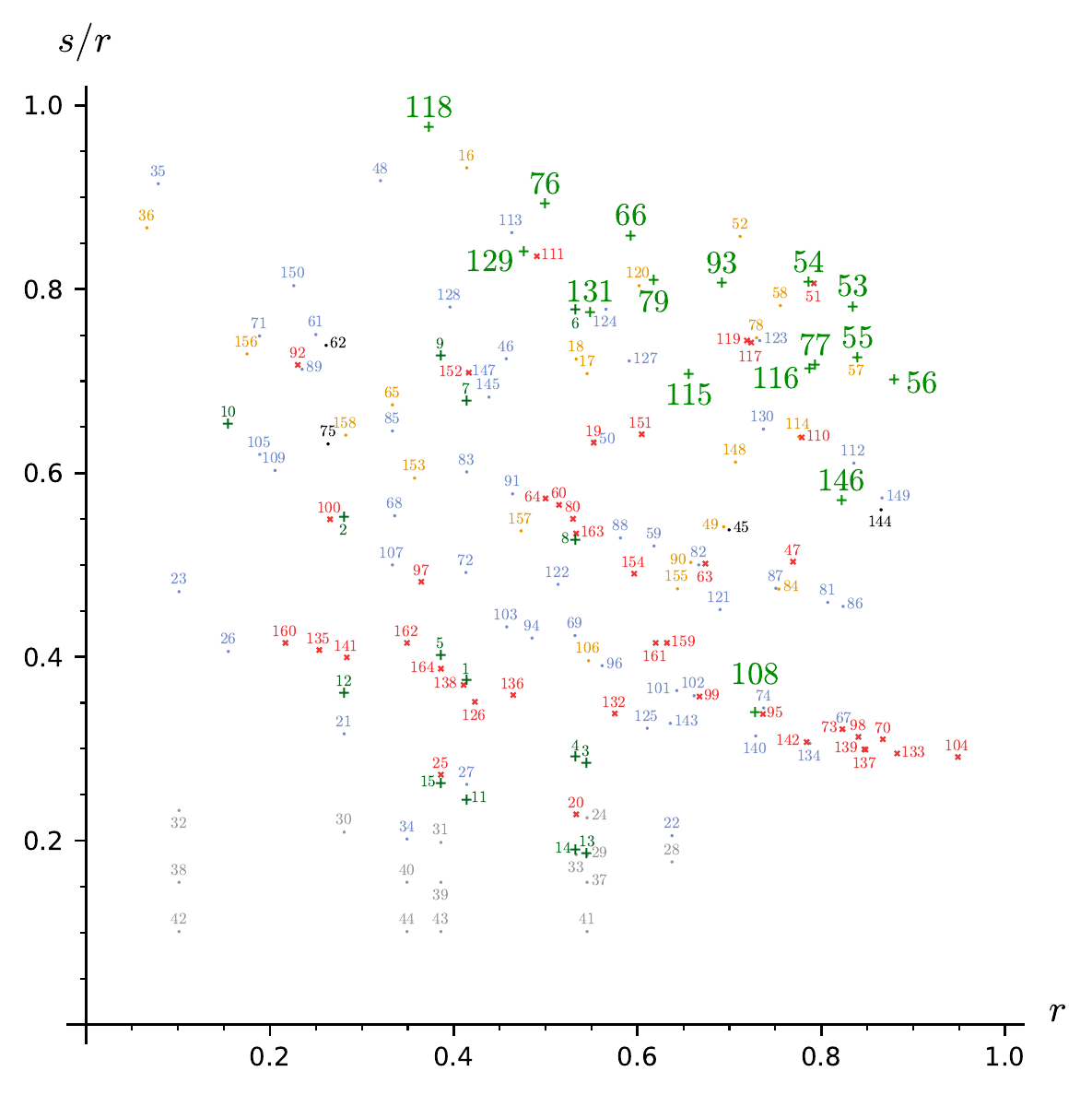}
  \caption{ The ``map'' of the 164 cases with triangulated ternary
    packings. The cases where no triangulated packing is saturated are
    marked in {\color{mapgrey} grey}. The cases with a ternary
    triangulated packing proved to maximize the density are marked by
    {\color{mapgreen} green} 
    \includegraphics[width=1.4ex]{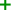}
    with larger case numbers.  The cases where we proved a
    triangulated binary packing to maximize the density are marked by
    {\color{mapdarkgreen} dark green} 
    \includegraphics[width=1.4ex]{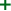}.
    The cases with
    counter examples are {\color{mapred} red}      
    (\includegraphics[width=1.2ex]{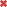}).
    The cases featuring two coronas (find the details in
    Section~\ref{sec:twocoronas}) are {\color{maporange} orange}. The
    cases with empty polytopes (see Section~\ref{sec:empty}) are
    {\color{mapblue} blue}. The remaining cases are marked in black
    (Section~\ref{sec:last}).}
  \label{fig:map}
\end{figure}

All in all, we proved the Connelly conjecture to be false and
classified the 149 cases where it was applicable in several groups: 16
cases for which the conjecture holds (Th.~\ref{thm:16}), 15 cases
where the density is maximized on a triangulated packing using only
two discs out of three (Th.~\ref{thm:15}), 40 (periodic) counter
examples to the initial conjecture (Th.~\ref{thm:40}), and the other
cases where our proof strategy does not work. Figure~\ref{fig:map}
represents each case as a point with coordinates $(r,\frac{s}{r})$ and
its number from~\cite{164}. The color of the point and the number
corresponds to the class we assigned to the case.

\contournumber{30}
\contourlength{0.5pt}

Section~\ref{sec:densest} is dedicated to the cases where a ternary
triangulated packing is proved to maximize the density. We start by
proving Th.~\ref{thm:16}
(Sections~\ref{sec:strat}--\ref{sec:construct_U}). We explain the
approach used in the similar proof for binary packings from \cite{BF}
and how we enhance it to make it work in our context. The first
improvement was the generalization of the code universal to all the
cases (instead of treating them one by one as in \cite{BF}). The
second necessary generalization we made was leaving a bunch of
parameters as free variables instead of fixing them arbitrary in the
beginning.

Our proof, as quite a few recent results in the domain, like~\cite{
  hales,fDiscs,ferniqueBounds}, is based on computer calculations. The
main details of the implementation are provided in
Section~\ref{sec:implement} (the complete code is given in the
supplementary materials). We prove Th.~\ref{thm:15} in
Section~\ref{sec:first18} by adjusting the proof of Th.~\ref{thm:16}.

Cases from Th.~\ref{thm:40} are treated in Section~\ref{sec:counter}. We
obtain a counter-example for each of these cases by applying the
flip-and-flow method~\cite{flipflow} on the triangulated binary
packings with disc radii ratio close to the radii ratios of pairs of
discs of this case.

Section~\ref{sec:others} is dedicated to the remaining cases.
Section~\ref{sec:twocoronas} presents the 22 cases where one of the
discs appears with at least two different neighborhoods. These cases
are analogous to the case b5 from Fig.~\ref{fig:9_triangulated} where
a small disc is either surrounded by 6 small discs or by two small
discs and two big ones. Our proof technique is not sufficient to treat
such cases. Handling them requires a less local approach, like it was
done in~\cite{BF} for b5.

Section~\ref{sec:empty} treats the 52 cases where we did not find a
set of constants satisfying all required inequalities needed in our
proof. Even though after several attempts with higher and higher
precision, we concluded that the existence of valid constants is quite
unlikely, it cannot be rigorously proved for the moment. We thus leave
this as an open problem.

Finally, Section~\ref{sec:last} is dedicated to the 4 cases where the
existence of such set of constants is more probable since we could
find the parameters satisfying the majority of constraints, but a few
of them were still not be satisfied. Whether the density is maximized
in these cases is also an open problem.

\section{When a triangulated packings is the densest}
\label{sec:densest}

In this section, we prove the first two parts of \cref{thm:1}.

\subsection{Proof strategy of Th.~\ref{thm:16}}\label{sec:strat}

We follow almost the same steps of the proof as in~\cite{BF} where the
same result is proven for binary triangulated packings and
in~\cite{thomas53} which treats computationally the ``simplest'' case
among the ternary triangulated packings (case 53). 

From the theoretical point of view, the transition from binary
packings to ternary ones seem to be straightforward. In practice,
however, we have much more cases to treat (149 instead of 9) and for
each of them, the problem is much more complex due to the high number
of local combinatorial configurations in possible packings. This
requires a more refined and sensitive choice of parameters than
in~\cite{BF}.

This section is strongly based on \cite{BF}: we use the idea of the
proof and quite a few intermediate results. Thus, for the sake of
simplicity, we preserve the same notations.

Let us describe the theoretical background of the proof which is
common for all cases, the only difference being the choice of the
parameters described in Section~\ref{sec:implement}.

We are given 3 discs of radii $1,r$ and $s$, $1{>}r{>}s$ and a ternary
triangulated packing of the plane by copies of these discs conjectured
to maximize the density, let us denote it by $P^*$. Our aim is to
prove that for any other packing $P$ using the same discs, its density
$\delta(P)$ does not exceed the density $\delta^*$ of $P^*$.

The main idea common to all the results about the maximal density of
triangulated packings was called ``cell balancing'' by
Heppes~\cite{heppes03} and it perfectly matches this title.  It
consists of two steps: first we locally ``redistribute'' the density
among some well-defined cells (triangles of the triangulation
in~\cite{BF,kennedy04,heppes03} and a mixture of Delaunay simplices
and Voronoi cells, both encoded in so-called decomposition stars,
in~\cite{HalesFormulation}) 
preserving the global density value. Then we prove that the
redistributed density of any cell of $P$ never exceeds $\delta^*$ .

First, let us define triangulations for packings by several sizes of
discs. The \emph{FM-triangulation} of a packing was introduced
in~\cite{fm} (it is a particular case of weighted Delaunay
triangulations~\cite{rourke}). Some of its useful properties are given
in~\cite{BF} (Section~4). The vertices of the FM-triangulation are the
disc centers. There is an edge between two disc centers iff there is a
point $p\in\RR^2$ and a distance $d>0$ such that $p$ is at distance
$d$ from the both discs and at least $d$ from any other
disc.

Let $\Tau$ and $\Tau^*$ respectively denote the FM-triangulations
of 
$P$ and $P^*$. The cells we are interested in are triangles of these
triangulations. Instead of working with densities, we introduce an
additive function $E$, called \emph{emptiness}, which, for a triangle
$T$ in $\Tau$, is defined by 
$$E(T):=area(T)\times\delta^* - area(T\cap P)\,.$$

This function was used in \cite{kennedy04} by the name of
``excess''. It was inspired by ``surplus area'' introduced
in~\cite{heppes03} defined as
$area(T) - \frac{area(T\cap P)}{\delta^*}$, identical to emptiness up
to multiplication by $\delta^*$. A similar but more complex function
called ``score'' is used in the proof of the Kepler
conjecture~\cite{lagarias02}.

The emptiness function reflects how ``empty'' the triangle is compared
to $\delta^*$. Indeed, $E(T)$ is positive if the density of $T$ is
less than $\delta^*$, negative if $T$ is denser, and equals zero if
$\delta(T)=\delta^*$. We use it rather than the density because of its
additivity: the emptiness of a union of two triangles equals the sum
of their emptiness values. This property does not hold for the
density.

To prove that $\delta\leq\delta^*$, it is enough to show that
$ \sum\limits_{T\in\Tau}E(T)\geq 0 $~\cite{BF}. This intuitively means
that $P$ is globally more empty and less dense than $P^*$.

Instead of working directly with the emptiness, we define a so-called
potential which plays the role of density redistribution mentioned
above. We do it since this function constructed explicitly is easier
to manipulate. We will construct a potential $U$ such that for any
triangle $T\in\Tau$, its potential does not exceed its emptiness:
  \begin{equation} 
    \label{ineq:tri}
    E(T)\geq U(T)  
  \end{equation}
  \noindent and the sum of potentials of all triangles in $\Tau$ is
  non negative:
  \begin{equation}
    \label{ineq:U}
    \sum_{T\in\Tau}U(T) \geq 0 
  \end{equation}\vspace{1ex}
  
  If, for $P^*$, there exists $U$ satisfying~(\ref{ineq:tri}) and
  (\ref{ineq:U}) for any packing $P$, then $P^*$ maximizes the density
  among packings using the same disc radii:
  
    $$(\text{\ref{ineq:tri}),(\ref{ineq:U})}\;\Longrightarrow\;\sum_{T\in\Tau}E(T)\geq 0 \; \Longrightarrow\; \delta^*\geq\delta\,.$$

    The rest of the proof consists in construction of potential $U$
    satisfying both (\ref{ineq:tri}) and (\ref{ineq:U}) for any
    packing $P$.

\subsection{How we choose the potential}
\label{sec:construct_U}
To construct the potential, we follow the idea first used by Kennedy
in~\cite{kennedy04} who introduced the ``localizing potential'' being
inspired by a statistical mechanics notion of
``m-potentials''. 
We define the total potential of a triangle as a sum of potentials of
``smaller'' units. It will consist of three \emph{vertex potentials}
$\Uv{}$ defined in the next section and three \emph{edge potentials}
$\Ue{}$ (Section~\ref{sec:edge}). Emptiness redistribution takes place
through vertex and edge potentials: the sum of vertex potentials
around each vertex in the triangulation will be non negative as well
as the sum of two edge potentials of triangles sharing this
edge. These two conditions guarantee us the ``global'' inequality
$(\ref{ineq:U})$.

All in all, the next four sections follow Sections~5.1-5.2,~6~and~7
of~\cite{BF}: we are constructing the potential in a way that the
global inequality $(\ref{ineq:U})$ is satisfied, at the same time
trying to minimize the potential of each triangle.  The last step is
to verify for any triangle that the local inequality
$(\ref{ineq:tri})$ is also satisfied, this part is discussed in
Section~\ref{sec:local}.

\subsubsection{Vertex potentials}
\label{sec:vertex}

We denote the first part of the potential by $\Uv{}$, it is composed
of three vertex potentials: if $A$, $B$, and $C$ are the vertices of
triangle $T$ in the FM-triangulation $\Tau$ of packing $P$,
$$\Uv{}(T):=\Uv{A}(T)+\Uv{B}(T)+\Uv{C}(T)\,.$$ Notice that we use the
same notations as in~\cite{BF} except that we add a dot above the
vertex potentials and a bar above the edge potentials to differentiate
them more easily.

We seek to choose the vertex potentials in a way to satisfy the
following inequality around any vertex $v\in\Tau$:
\begin{equation}
  \label{eq:ver}
  \sum_{T\in\Tau |v \in T}\Uv{v}(T) \geq 0\tag{$\bullet$}\,.
\end{equation}

\begin{figure}
  \centering
  \includegraphics[height=16ex]{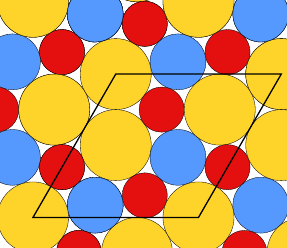}\hskip4ex
  \includegraphics[height=16ex]{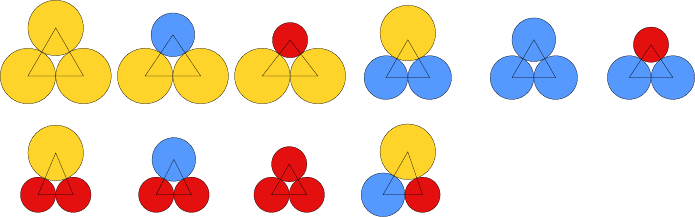}
  \caption{Case 54, its triangulated packing (left) and its 10 tight
    triangles (right).}
  \label{fig:tight_triangles}
\end{figure}

\noindent This inequality implies global inequality~(\ref{ineq:U}) for vertex
potentials.

A triangle is called \emph{tight} if it is formed by three pairwise
tangent discs. It is known that the properties of the FM-triangulation
imply that all triangles of a triangulated packing are tight. For each
set of three distinct disc radii, there are 10 tight triangles, one
for each triplet of discs (see Figure~\ref{fig:tight_triangles}). Let
$E_{xyz}$ denote the emptiness of the tight triangle formed by discs
of radii $x,y,z$. We denote by $V_{xyz}$ the vertex potential of this
triangle in the vertex corresponding to the $y$-disc. Since we set
$V_{xyz}=V_{zyx}$, there are 18 of them.

\begin{figure}[ht]
  \centering
  \includegraphics[height=13ex]{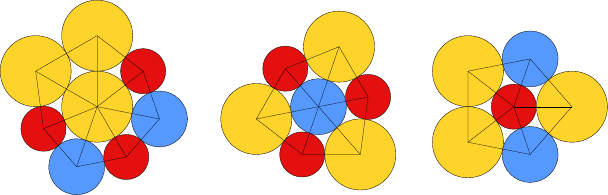}
  \caption{Three coronas of packing 54.}
  \label{fig:coronas}
\end{figure}\vskip1ex

Inequalities $(\ref{ineq:tri})$ and $(\ref{ineq:U})$, applied to
$\Tau^*$, imply the following constraints on the vertex potentials of
tight triangles appearing in $\Tau^*$. 
For each tight triangle $T$ in $\Tau^*$, $\Uv{}(T)=E(T)$ and for each
vertex $v$ in $\Tau^*$,
\begin{equation}
  \label{eq:tri_cor}
  \sum\limits_{T\in\Tau^*|v\in T}{\Uv{v}(T)}=0.
\end{equation}
The latest means that
the sum of vertex potentials in $v$ of all triangles surrounding it
(the set of these triangles is called the \emph{corona} of $v$) equals
zero. Let us illustrate it with the case $54$ whose triangulated
packing is given in Fig.~\ref{fig:tight_triangles}. There are only
three types of tight triangles present in this packing, so the first
class of constraints produces the following equations:
$$3V_{111}=E_{111},\qquad V_{1s1}+2V_{11s}=E_{1s1},\qquad  V_{1rs}+V_{rs1}+V_{s1r}=E_{1rs}\,. $$
The equations for the three coronas given in Fig.~\ref{fig:coronas}
are respectively:
$$V_{111}+2V_{11s}+4V_{r1s}=0,\qquad 6V_{1rs}=0,\qquad  V_{1s1}+4V_{1sr}=0\,.$$

We thus have $18$ variables and $6$ equations with only $5$ of them
being independent due to periodicity of the packing: since the sum of
the emptiness of tight triangles of the fundamental domain of the
packing equals zero, there is a linear combination of tight triangle
potentials equal to zero.

We also set the potentials of all other tight triangles equal to their
emptiness. In our example, this gives us $7$ more equations that are
still independent and leaves us with $6$ free variables.

Until now we followed the strategy used in \cite{BF} which does not
work for the cases proved here, except the case $53$ proved earlier
in~\cite{thomas53}. The problem is that in~\cite{BF,thomas53}, the
free variables mentioned above are all directly set to zero to
simplify further computations. The first generalization we made was to
keep these $6$ variables as free variables and fix them later, in
order to satisfy all inequalities $(\ref{ineq:tri})$ around the
vertices. The $6$ tight vertex potentials left free are the vertex
potentials of the isosceles (but not equilateral) tight triangles in
the vertex adjacent to the two equal edges. More formally, the free
variables are the following:
$$\{V_{xyx}\;|\;x,y\in\{1,r,s\},x\neq y\}.$$ These variables are always
independent and independent of the equations.

The next step is to choose the vertex potentials of all the other (non
tight) triangles in a way that inequalities (\ref{ineq:tri}) and
(\ref{ineq:U}) hold.

As in Section~5.2 of~\cite{BF}, we define them as follows:
$$\Uv{v}(T):=V_{xyz}+m_y|\hat{v}-\widehat{xyz}|,$$

\noindent where $x,y,z$ are the disc sizes of $T$, $v$ is the vertex
corresponding to the $y$-disc, $\hat{v}$ is the angle of $T$ in $v$,
$\widehat{xyz}$ is the angle in the vertex of the $y$-disc in the
tight triangle with discs of radii $x,y,z$, and $m_y$ is a constant
defined below. The difference between the vertex potential of $T$ and
the vertex potential of the tight triangle with the same discs is
proportional to their angle difference, or ``how different triangle
$T$ is from the tight one''. The constants $m_1,m_r,m_s$ reflecting
the ``importance'' of this angle deviation in vertex potentials should
be chosen carefully, which is explained below.

Given a triangle $T$ with discs of radii $x,y,z$ in $\Tau$, let $T^*$
denote the tight triangle formed by the same discs.

Our aim now is to choose the tight vertex potentials and $m_1,m_r,m_s$
in a way that inequality (\ref{eq:ver}) holds around each vertex.
This is the case if, for each vertex $v$ with disc of radius $q$ in
$\Tau$, the following inequality is satisfied:

\begin{equation}
  \label{eq:mq}
  m_q \geq \frac{-\sum\limits_{j=1}^k{\Uv{}(T^*_j)}}{\left|2\pi - \sum\limits_{j=1}^k{\widehat{v}(T^*_j)}\right|}\,,
\end{equation}

\noindent where $T_1,\dots,T_{k}$ is the corona of $v$ in $\Tau$.

The proof is identical to the one for binary packings given
in~\cite{BF}. Notice that the only case where the denominator equals
zero is when $T^*_1,\dots,T^*_{k}$ form a corona in the triangulated
packing. The sum of vertex potentials of triangles in such corona
equals zero by~(\ref{eq:tri_cor}).

We thus have to choose the tight vertex potentials and $m_1,m_r,m_s$
satisfying these inequalities. As explained in the proof of
Proposition~3 of \cite{BF}, it is enough to perform an exhaustive
search on a finite number of configurations to assure this
inequality. It thus can be done by a computer.

The $6$ values of potentials $V_{xyz}$ of tight triangles which were
left as free variables, as well as $m_1,m_r,m_s$, are chosen to
satisfy inequality~(\ref{eq:mq}). The solutions of this inequality (a
subset of $\RR^9$) are the appropriate combinations of values of
potentials and $m_1,m_r,m_s$. The details of computer implementation
of this part and how we choose the precise values are described in
Section~\ref{sec:poly}. 

Furthermore, using Proposition~4 from Section~6 of~\cite{BF}, we can
cap each vertex potential with a constant value depending only on the
disc type of the vertex. The proof of this proposition roughly
consists in showing that for any vertex, as soon as the vertex
potential of a triangle in its corona exceeds a certain value, the
remaining vertex potentials will never be ``too negative'' which
implies inequality~(\ref{eq:ver}) in this vertex.

This allows us to diminish vertex potential
while still satisfying its non negativity around each vertex
(\ref{eq:ver}). It was not needed in~\cite{thomas53} for case 53 but
turns out to be necessary for all the other cases we consider. Thus,
the vertex potential of triangle $T$ in vertex $v$ corresponding to a
$q$-disc will be rewritten as $Z(\Uv{v}(T))=\min{(\Uv{v}(T), Z_q)}$
where

$$Z_q:=2\pi\,\left|\min\limits_{T_{xqz}\in \Tau}{\frac{V_{xqz}}{\widehat{xqz}}}\right|.$$

\subsubsection{Edge potentials}
\label{sec:edge}

As in Section~7 of~\cite{BF}, in order to keep the potential lower
than the emptiness in all triangles (to satisfy
inequality~(\ref{ineq:tri})), we introduce its second part, $\Ue{}$,
called the edge potential. We can now write down the total potential
$U(T)$ of a triangle $T$:
$$U(T):=Z(\Uv{}(T))+\Ue{}(T)\,.$$

We need $\Ue{}$ to compensate the vertex potential of ``stretched''
triangles: those where one angle is too large. Such triangles feature
high vertex potential and low emptiness. The edge potential of a
triangle is equal to the sum of edge potentials corresponding to its
three edges: if $e,d,f$ are the edges of triangle $T$,
$\Ue{}(T):=\Ue{e}(T)+\Uv{d}(T)+\Uv{f}(T)$.

We define the edge potential of $T$ in $e$ as
follows:\\ 
\begin{equation*}
  \Ue{e}(T) := 
  \begin{cases}
    q_{xy} d_e  & \text{if } |e|>l_{xy} \\
    0 & \text{otherwise}
  \end{cases}
\end{equation*}
\noindent where $x,y$ are the radii of the discs with centers in the endpoins of
edge $e$ and $d_e$ is the signed distance from the center $X$ of the
circumscribed circle of $T$ to $e$ ($d_e$ is positive iff $X$ and $T$
are at the same side with respect to $e$). The choice of the constants
$q_{xy},l_{xy}$ is explained below.

With this definition, the sum of edge potentials of two triangles
sharing this edge is always non negative. This is shown in \cite{BF}
(Section~7, Proposition~3). Thus, adding the edge potential keeps the
global inequality (\ref{ineq:U}) valid. Meanwhile, it affects the
local distribution of potentials among neighbour triangles approaching
us to our aim which is to satisfy $E(T)\geq U(T)$ in all triangles.

We pick the pairs of constants $q_{xy}$ and $l_{xy}$ (there are 6 of
them, one for each pair $(x,y)$ of disc radii) in order to compensate
high vertex potential of stretched triangles. To find constants
allowing to do this, we compute the vertex potential and the emptiness
of the most ``dangerous'' triangles: those with two pairs of tangent
discs (see an example in Figure~\ref{fig:curves}). We represent them
as curves in function of the length of the only ``flexible'' edge.

\definecolor{db}{RGB}{0,0,128}
\definecolor{dg}{RGB}{0,95,54}
\definecolor{ro}{RGB}{219,99,255}

\begin{figure}[htpb]\vspace{-25ex}
  \centering
  \includegraphics[width=.75\linewidth]{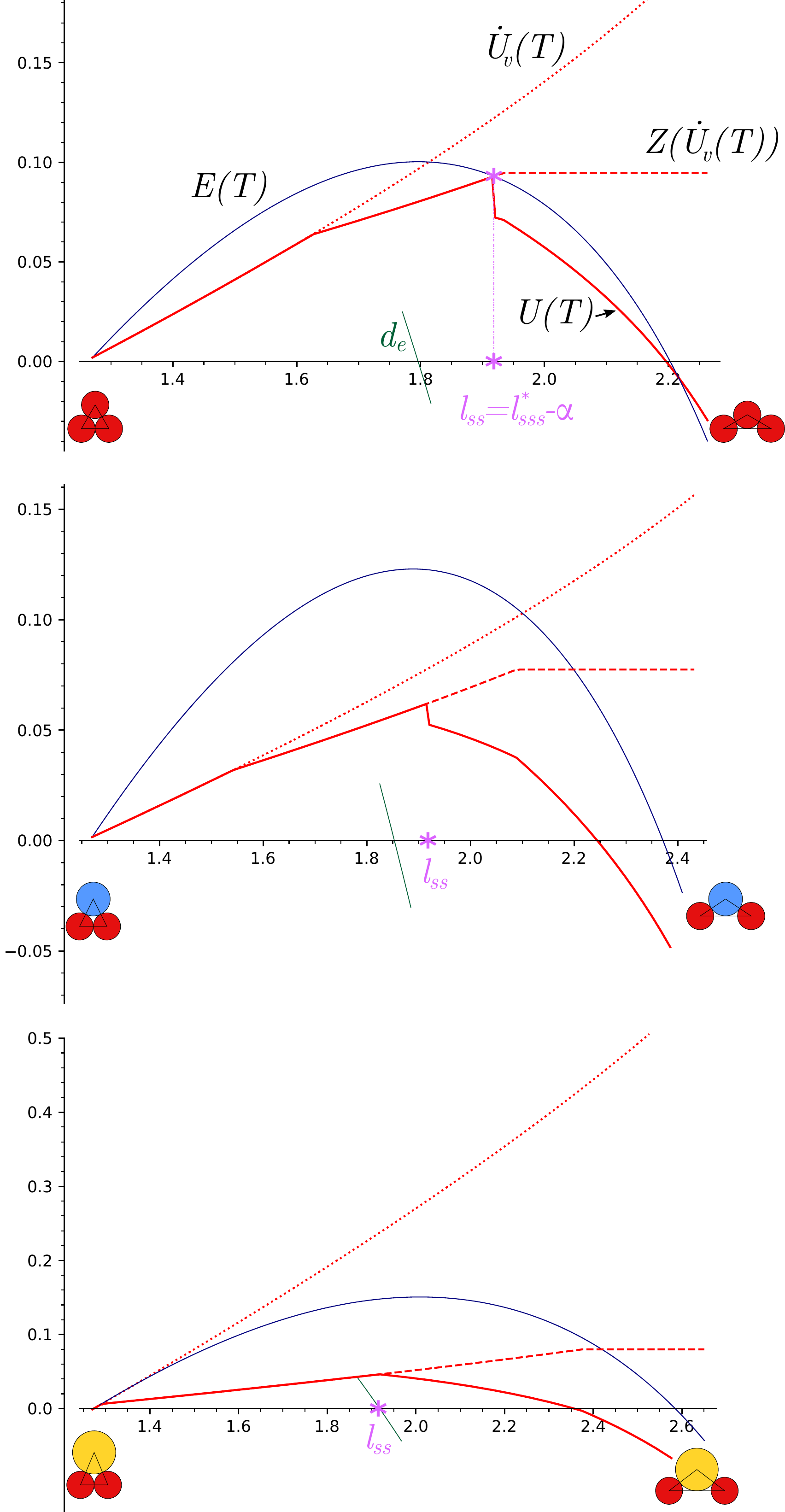}
  \caption{Case 54, triangles with discs $sss$, $srs$, $s1s$: behavior
    of the emptiness (in {\color{db} dark blue}) and the potential
    while stretching an edge. Initial vertex potential $\Uv{v}(T)$ is marked
    by the {\color{red}dotted red line}, capped vertex $Z(\Uv{v}(T))$
    potential is the {\color{red}dashed red line}, the total potential
    $U(T)$ is the \textbf{\color{red}bold red line}. The {\color{dg}
      dark green segment} designates $d_e$ around the moment it
    becomes negative, and the {\color{ro} pink asterisk}
    \includegraphics[height=6pt]{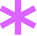} indicates the value of
    $l_{ss}$.}
  \label{fig:curves}
\end{figure}

To choose $q_{xy}$ and $l_{xy}$ for given $(x,y)$, we consider three
triplets of discs radii: $x1y, xry,xsy$ and for all of them we trace
the curves as ones shown in Figure~\ref{fig:curves}. The aim of edge
potential is not to let the capped vertex potential $Z(\Uv{v}(T))$
({\color{red}dashed red line}) exceed the emptiness $E(T)$
({\color{db}dark blue line}). In all the cases considered here, the capped
vertex potential and the emptiness have at most one intersection
except the leftmost point corresponding to the tight triangle (the
neighborhood of this point is a special case treated in detail in
Section~\ref{sec:epstight}). This intersection is the side length such
that stretching the triangle even more causes vertex potential to be
greater than emptiness.

Let $l^*_{xqy}$ be the side length where this intersection occurs for
a triangle formed by discs with radii $x,q,y$. Notice that $d_e$ is a
decreasing monotonous function on the side length $|e|$. If
$d_{l^*_{xqy}}<0$ (which is the case for all the proved cases), then
it is enough to set $l_{xy}$ equal to at most $l^*_{xqy}-\alpha$ with
a small enough $\alpha$ (we used $\alpha=10^{-5}$). We thus set
$$l_{xy}:=\min\limits_{q=1,r,s}l^*_{xqy}-\alpha.$$
As an illustration, when choosing $q_{ss}$ and $l_{ss}$ for case 54
(see Fig.~\ref{fig:curves}), the triangle with discs of radii $s,s,s$
has the leftmost intersection of $E(T)$ and $Z(\Uv{v}(T))$
(\includegraphics[height=6pt]{pictures/ast} on the first graph). This
means, $l_{ss}=l^*_{sss}-\alpha$.

Then we choose the coefficient $q_{xy}$ in a way that
$Z(\Uv{v}(T)) + d_eq_{xy}$ stays below $E$ starting from
$l^*_{xqy}-\alpha$ for all $q=1,r,s$ (which is always possible since $d_e$
is negative on this
segment).

These choices guarantee the total potential $U(T)$
(\textbf{\color{red}bold red line} in Fig.~\ref{fig:curves}) to be
below the emptiness all the way from tight to stretched triangles.

\subsection{Verifying local inequality (\ref{ineq:tri})}
\label{sec:local}

In Section~\ref{sec:construct_U}, we constructed the potential in a
way that the global inequality $(\ref{ineq:U})$ is satisfied. We
should now prove that for any triangle $T$, (\ref{ineq:tri}) holds:
$E(T)\geq U(T)$.  This part widely uses computer calculations: we execute
them in SageMath~\footnote{\url{https://www.sagemath.org/}}.  We first treat
the triangles that are close to the tight ones, so-called
\emph{epsilon-tight triangles}, in Section~\ref{sec:epstight}. For all
the other triangles, (\ref{ineq:tri}) is verified in
Section~\ref{sec:alltriangles}.

\subsubsection{Epsilon-tight triangles}
\label{sec:epstight}
As in Section~8 of~\cite{BF}, the definition of the vertex potential
allows us to directly show that it does not exceed the emptiness in
the triangles ``close enough'' to the tight ones.  We do it by
comparing the differentials of the two functions. We are interested in
this case since tight triangles are the limiting point where the
emptiness is equal to the potential.

Let $\Tau_{\epsilon}$ denote the set of \emph{epsilon-tight}
triangles, those where each pair of discs is at distance at most
$\epsilon$. We can bound the variation $\Delta E$ of emptiness on
$\Tau_{\epsilon}$:

$$\Delta E \geq \sum\limits_{i=1,2,3}\min_{\Tau_\epsilon}{\frac{\partial E}{\partial x_i}\Delta x_i}, $$

\noindent where $x_1,x_2,x_3$ are the lengths of the sides.

We consider only the values of $\epsilon$ that are strictly smaller
than the $l_{xy}-$constants so that edge potentials do not play a role
in this case: since the smallest $l_{xy}$ is greater than $\epsilon$,
we can find the lower bound of the variation $\Delta U$ of potential:

$$\Delta U \leq \sum\limits_{i=1,2,3}\max_{\Tau_\epsilon}\frac{\partial U}{\partial x_i}\Delta x_i\,.$$

As we have set the potential equal to the emptiness on tight
triangles, choosing $\epsilon$ such that for $i=1,2,3$,

\begin{equation}
  \label{ineq:eps}
  \min_{\Tau_\epsilon}{\frac{\partial E}{\partial x_i}\Delta x_i} \geq \max_{\Tau_\epsilon}\frac{\partial U}{\partial x_i}\Delta x_i
\end{equation}
guarantees us that the local inequality~(\ref{ineq:tri}) is satisfied
by all triangles in $\Tau_\epsilon$.

To find the greatest value of $\epsilon$ satisfying this inequality,
we use a dichotomic search, aiming to maximize $\epsilon$ (indeed,
maximizing the set $\Tau_\epsilon$ we minimize the set of the
remaining triangles that should be treated afterwards). At each step
of the binary search, we compute both values in interval arithmetic
(explained in the next section and Section~\ref{sec:rif}) and if the
intervals of these values do not intersect and satisfy
inequality~(\ref{ineq:eps}), the given $\epsilon$ is valid.

\subsubsection{Other triangles}
\label{sec:alltriangles}

The final step is to verify the local inequality $(\ref{ineq:tri})$ on
the set of non epsilon-tight triangles that could appear in a
triangulation of a packing.  Thanks to the properties of
FM-triangulations of saturated packings, this set is compact. Indeed,
the length of the edge connecting centers of discs of radii $x$ and
$y$ in a triangle is at least $x+y$ and at most $x+y+2s$. The
\emph{support circle} of a triangle in a packing is the circle tangent
to the three discs with centers in the vertices of the triangle.
Thanks to FM-triangulation properties, the support circle never
intersects other discs of the triangulation. The upper bound on the
side length comes from the fact that the radius of the \emph{support
  circle} is at most s (otherwise the packing would not be saturated).
Thus, if the centers of the two discs are respectively $X$ and $Y$ and
the support circle center is $O$, applying the triangle inequality to
$XOY$, we get $|XY|\leq x+y+2s$.

Working with intervals rather than precise values allows us to
decompose a compact continuum set of cases into a finite one.  Given a
triplet of disc radii, each side of each triangle with these 3 discs
in vertices is bounded. Therefore, to prove inequality $E(T)\geq U(T)$
for all triangles with given disc radii, it is enough to show it for
the triangle with sides represented by intervals depending on these
radii. More precisely, we should verify the following inequality:
$$E(\overline{T}_{xyz}) \geq U(\overline{T}_{xyz}),$$
\noindent where $\overline{T}_{xyz}$ is a triangle with discs of radii
$x,y,z$ and sides represented by intervals
$[y+z, y+z+2s],[x+z, x+z+2s],[x+y,x+y+2s]$. If it holds (the returned
value is \texttt{True}), then the inequality holds for all possible
triangles with discs $x,y,z$. If the returned value is \texttt{False},
this means either that the inequality is false or that the intervals
in question intersect (and thus are incomparable). In this case, we
subdivide initial intervals augmenting
precision. Section~\ref{sec:rif} describes in detail how it is done
in practice, using interval arithmetic tools of SageMath.

\subsection{Computer implementation}
\label{sec:implement}

As many proofs of the domain, notably the proof of the Kepler
Conjecture~\cite{hales}, the proofs of the maximal density for
triangulated packings, like ours and those
from~\cite{BF,thomas53,kennedy04}, essentially rely on computer
calculations. In this section, we discuss the details of computer
implementation.

The treatment of each case consists of two steps. We first choose all
the values necessary to define the potential: tight vertex potentials
$V_{xqy}$, constants $m_q$ and capping values $Z_q$ from
Section~\ref{sec:vertex}, the value of $\epsilon$
(Section~\ref{sec:epstight}), and the constants $l_{xy},q_{xy}$ of the
edge potentials (Section~\ref{sec:edge}). We choose them in a way to
satisfy the ``global'' inequality~(\ref{ineq:U}). The second step is to
verify the ``local'' inequality~(\ref{ineq:tri}) on all possible
triangles explained in Section~\ref{sec:alltriangles}.

\subsubsection{Interval arithmetic}
\label{sec:rif}
We use interval arithmetic in two completely different contexts: to
work with real numbers non representable in computer memory and to
verify inequalities on uncountable but compact sets of values. More
precisely, we use intervals to store the values of radii of discs
which are algebraic numbers obtained as roots of polynomials
in~\cite{164} as well as the value of $\pi$. The other situation where
we use intervals is to verify the local inequalities on a compact
continuum set of triangles in Section~\ref{sec:alltriangles}.

In interval arithmetic, each value is represented by an interval
containing this value and whose endpoints are exact values finitely
representable in computer memory (floating-point numbers). Performing
functions in interval arithmetic preserves both properties. More
precisely, if $x_1,\dots,x_n$ are intervals, and $f$ is an $n$-ary
function, the interval $f(x_1,\dots,x_n)$ must contain
$f(y_1,\dots,y_n)$ for all
$(y_1,\dots,y_n)\in x_1{\times}\dots{\times} x_n$ and its endpoints are
floating-point numbers.

To verify an inequality on two intervals $x_1 < x_2$, it is enough to
compare the right endpoint of $x_1$ and the left endpoint of
$x_2$. The returned value is \texttt{True} only if each pair of values
from these intervals satisfy the inequality. However, if the result is
\texttt{False}, that does not mean that the inequality is false on the
numbers represented by $x_1$ and $x_2$, it might also mean that these
intervals overlap.

We worked with interval arithmetic implemented in
SageMath~\cite{sage}, called Arbitrary Precision Real
Intervals~\footnote{\url{https://doc.sagemath.org/html/en/reference/rings_numerical/sage/rings/real_mpfi.html}}. The
intervals endpoints are floating-point numbers, the precision we use
in the majority of cases is the default precision of the library where
the mantissa encoding has 53 bits.

\subsubsection{Polyhedra}
\label{sec:poly}

In Section~\ref{sec:vertex}, we choose the values of vertex potentials
in tight triangles and constants $m_1,m_r,m_s$ in a way to satisfy all
the necessary constraints. These constraints together define a subset
of $\RR^9$ (where the variables are 6 tight vertex potentials
$V_{1r1},V_{1s1},V_{r1r},V_{rsr},V_{s1s},V_{srs}$ and 3 constants
$m_1,m_r,m_s$). We use the Polyhedra
module\footnote{\url{https://doc.sagemath.org/html/en/reference/discrete_geometry/sage/geometry/polyhedron/constructor.html}}
of SageMath to work with them (it allows us to store the solutions of
a system of linear inequalities as a convex polyhedron).

Even more constraints are added in Section~\ref{sec:epstight}, since
there should exist a positive value of $\epsilon$ satisfying the
inequalities~(\ref{ineq:eps}). To guarantee that, we verify if
inequality~(\ref{ineq:eps}) holds for \texttt{$\epsilon=0$}, in other
words, we make sure that this inequality holds for some non-negative
$\epsilon$. We do it in SageMath: to compute both parts of the
inequality, we use interval arithmetic and calculations of
derivatives. The obtained inequalities are intersected with the
polyhedron calculated above. For all the cases considered in this
section, this intersection is not empty (the cases where it was empty
are discussed in Section~\ref{sec:empty}). Then we find the maximal
value of $\epsilon>0$ allowing the intersection not to be empty and
this permits us to fix $\epsilon$.

For all the cases we treat in this section, these constraints together
define a compact polyhedron in $\RR^9$ (where the variables are the 6
tight vertex potentials and $m_1,m_r,m_s$).


After we get a polyhedron of valid values, we are free to choose a
point inside to fix them. Our aim at that step is to minimize
potentials of all triangles in order to satisfy~(\ref{ineq:tri}). We
thus find the three vertices of the polyhedron minimizing $m_1$, $m_r$
and $m_s$ respectively, compute a linear combination of them (the
weights that worked well in practice were respectively 1,1 and
4), 
and take a point between this one and the center of the polyhedron in
order to avoid the approximations problems on the border which are
discussed in the next paragraph. Our method to choose the point
described above is a heuristic.

Implementing construction of polyhedra, we encounter the following
problem: the Polyhedra class does not allow coefficients of
constraints to be intervals, while some of the coefficients of our
inequalities are stored as such due to their dependency of $\pi$ and
disc radii. Polyhedra do not support intervals as a base ring for a
good reason: solutions of a system of linear inequalities with
interval coefficients might not form a convex polyhedron. We choose to
replace the intervals with their centers and work with an
approximation of the actual set of valid values for tight potentials
and $m_1,m_r,m_s$. Our polyhedron is stored in a field of rational
values, since this field is computationally quite efficient.

That means, after choosing a point inside this approximated polyhedra,
we can not know if this point actually satisfies all the
constraints. To make sure it does, we then rigorously verify that all
the inequalities with interval coefficients hold in this point.

\subsection{Proof of Th.~\ref{thm:15}}
\label{sec:first18}
\begin{figure}[htp]
  \center
  \includegraphics[width=\textwidth]{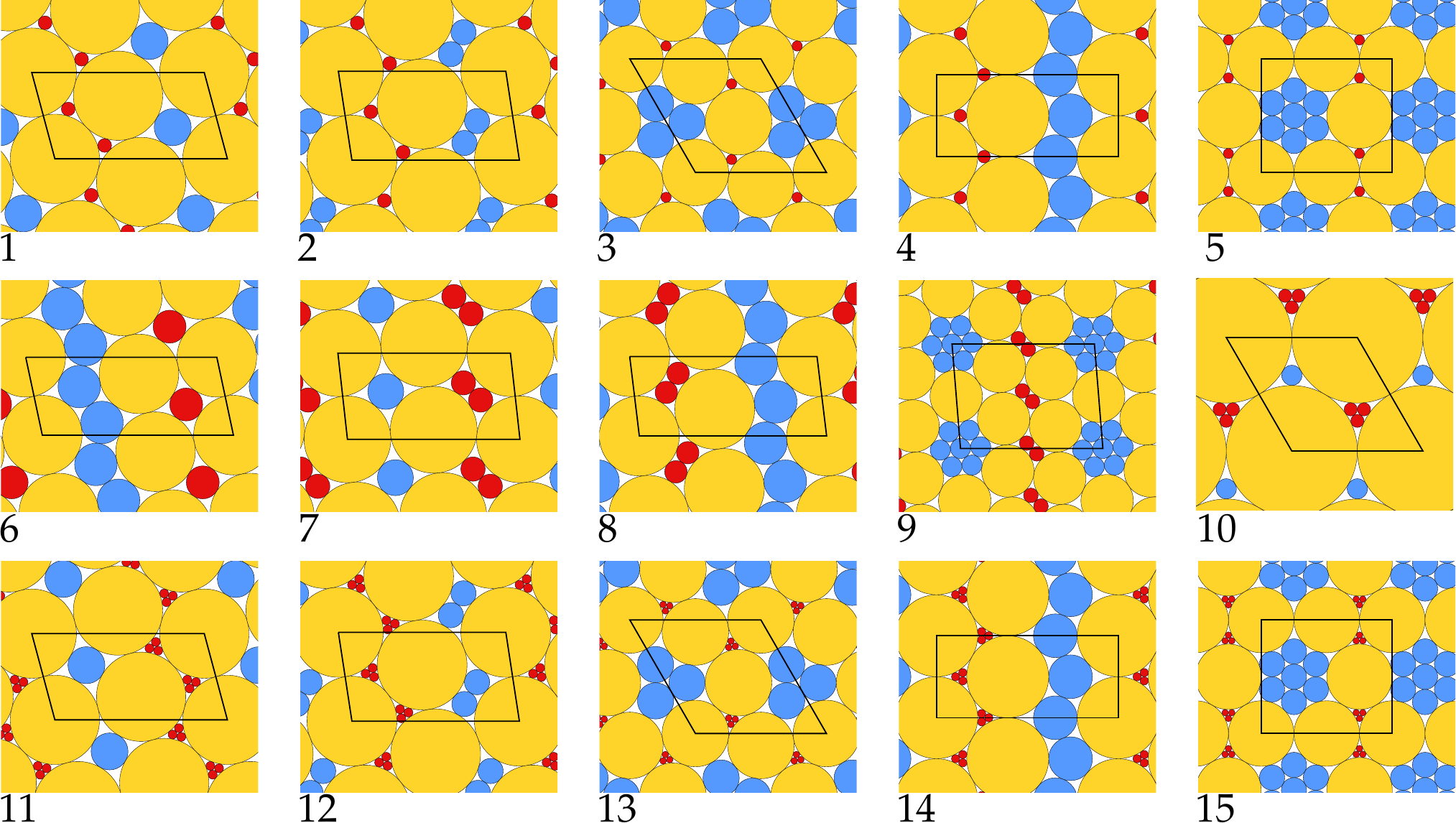}
  \caption{Triangulated ternary packings for cases 1-15, where a
    triangulated binary packing maximizes the density. For cases 1--5,
    it is the triangulated packing of $b_8$; for case 6 --- $b_4$; for
    cases 7--9 --- $b_7$; for cases 10--16 --- $b_9$. }
  \label{fig:1-15}
\end{figure}

Cases 1-18 are special: they are called \emph{large separated}
in~\cite{164} since they do not contain pairs of adjacent medium and
small discs (see Fig.~\ref{fig:1-15} for the first 15). For each of
these cases, in addition to ternary triangulated packings, there are
other triangulated packings using only two discs out of three. It
happens because the radii of small and medium discs coincide with the
radii of small discs of two cases among $b_1$-$b_9$. It is thus
possible to assemble packings having the same density as the binary
packings of mentioned cases using only two of three discs. It turns
out that in all these cases, the density of one of the mentioned
binary packings exceeds the density of the ternary one. That means,
for each of cases 1-18, the densest packing among the triangulated
ones is a binary packing corresponding to a case from $b_1$-$b_9$
(Fig.~\ref{fig:9_triangulated}).

\vskip1ex 

Indeed, each of these ternary packings is formed as a ``combination''
of two binary packings one of which is denser than the other. Thus,
the densest of the binary packings will also be denser than its
combination with a less dense packing.

We were able to show that the denser triangulated binary packing
maximizes the density among all packings (not only triangulated ones)
for the cases from 1 to 15 (Fig.~\ref{fig:1-15}). The proof is almost
the same as in Section~\ref{sec:densest}. 

Let $i$ be the case number and $P_3$ denote its triangulated ternary
packing. Let $P^*_2$ denote the densest triangulated binary packing
using two discs of case $i$ and let $P_2$ denote the less dense
triangulated binary packing using two discs of case $i$. We already
know that $P^*_2$ is denser than the two others,
$\delta(P^*_2)>\delta(P_3)>\delta(P_2)$. Our aim is to show that
$P^*_2$ maximizes the density among all packings by the discs of case
$i$.

The only difference with the strategy used for other cases concerns
vertex potentials from Section~\ref{sec:vertex}. Since $P^*_2$ uses
only two discs out of three, it features only 2 coronas instead of
3. Thus, these 2 coronas together with the 10 equations for tight
triangles, give us at most 11 independent equations instead of 12.

We now need to chose 7 free variables instead of 6. We can pick 6
tight potentials of isosceles triangles as before. There remains to
choose the last free variable. Vertex potentials of equilateral tight
triangles can not be picked because of the equations of type
$V_{xxx}=E_{xxx}$: they are already fixed. The remaining vertex
potentials of isosceles triangles ($V_{xxy},\; x\neq y$) can not be
used since they are dependent of the first 6 free variables and the
equations $2V_{xxy}+V_{xyx}=E_{xyx}$. The only candidates thus are
$V_{1rs},V_{1sr},V_{r1s}$; we add one of them.

For cases 16, 17, and 18, the densest binary packing is $b_5$ which
features two different coronas around the small disc, so our method is
not applicable to them as discussed in Section~\ref{sec:twocoronas}.

To summarize, for cases 1-18, among triangulated packings, the density is
maximized by a binary packing, not a ternary one as in the Connelly
conjecture. However, whether this packing maximizes the density among
all packings is still an open question for cases 16, 17 and 18.

\section{Counter-examples: proof of Th.~\ref{thm:40}}
\label{sec:counter}

Starting to work on the density of ternary saturated triangulated
packings, we believed the Connelly conjecture to hold, i.e. that for
all of the 149 cases, a triangulated packing would maximize the
density. Realization that our proof strategy failed for many of them
made us suspect the conjecture to be false. Knowing that the density
of binary triangulated packings (all of them are given in
Figure~\ref{fig:9_triangulated}) often exceeds the density of ternary
triangulated packings in question gave us an idea to use them in order
to find counter examples.

The first result we obtained was for case 110~\cite{counter}. After
generalization, we ended up with 40 counter examples (19, 20, 25, 47,
51, 60, 63, 64, 70, 73, 80, 92, 95, 97, 98, 99, 100, 104, 110, 111,
117, 119, 126, 132, 133, 135, 136, 137, 138, 139, 141, 142, 151, 152,
154, 159, 161, 162, 163, 164). They are all non triangulated packings
using only two discs out of three which have greater densities than
triangulated packings using all three discs. We obtained each of them
deforming a triangulated binary packing with discs whose size ratio is
close to the one of a pair of discs in the triplet associated to the
case. Tiny deformations do not dramatically lower the density and
these packings are dense enough to outplay the ternary triangulated
ones.

\begin{figure}[h]
  \center
  \includegraphics[width=.32\textwidth]{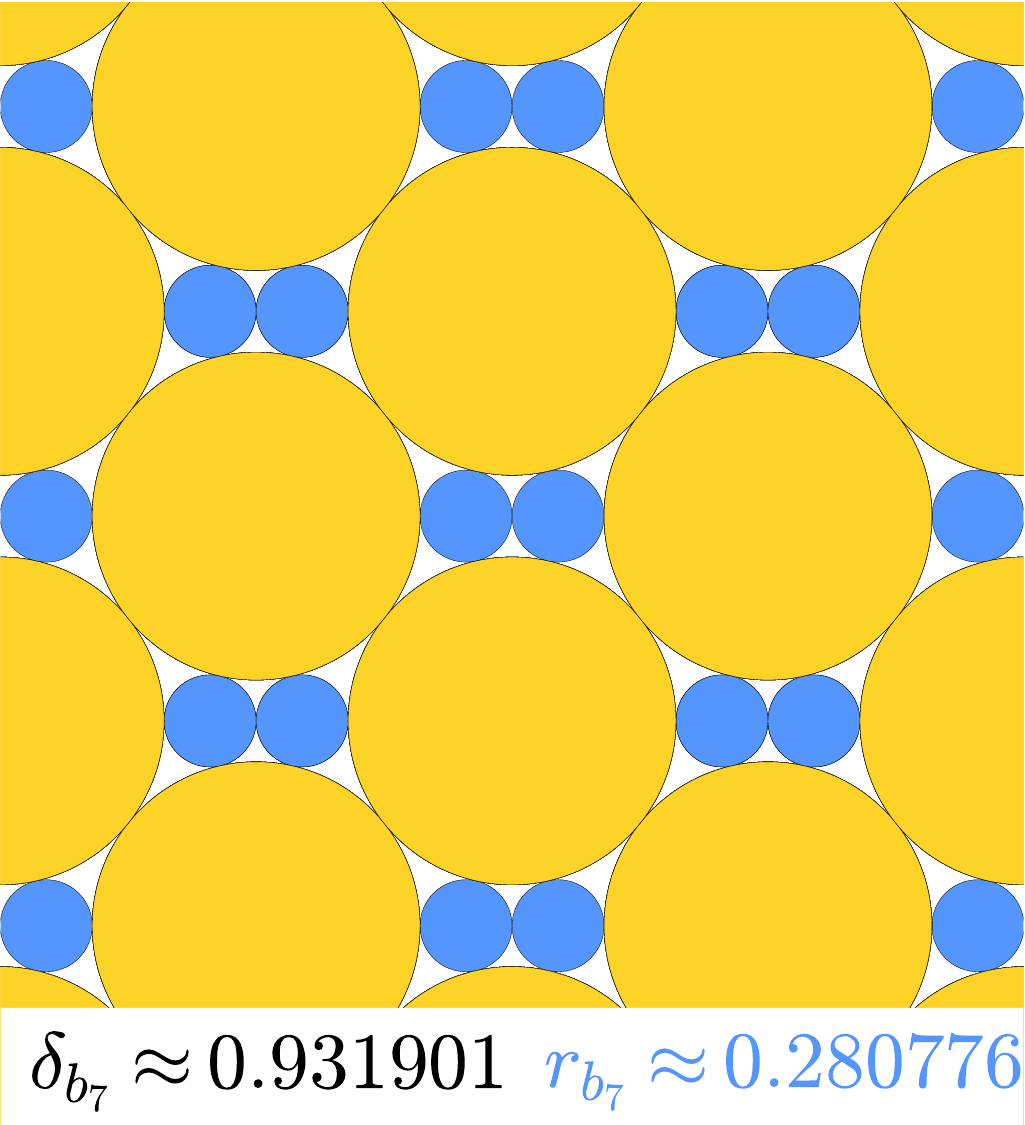}\hfill
  \includegraphics[width=.32\textwidth]{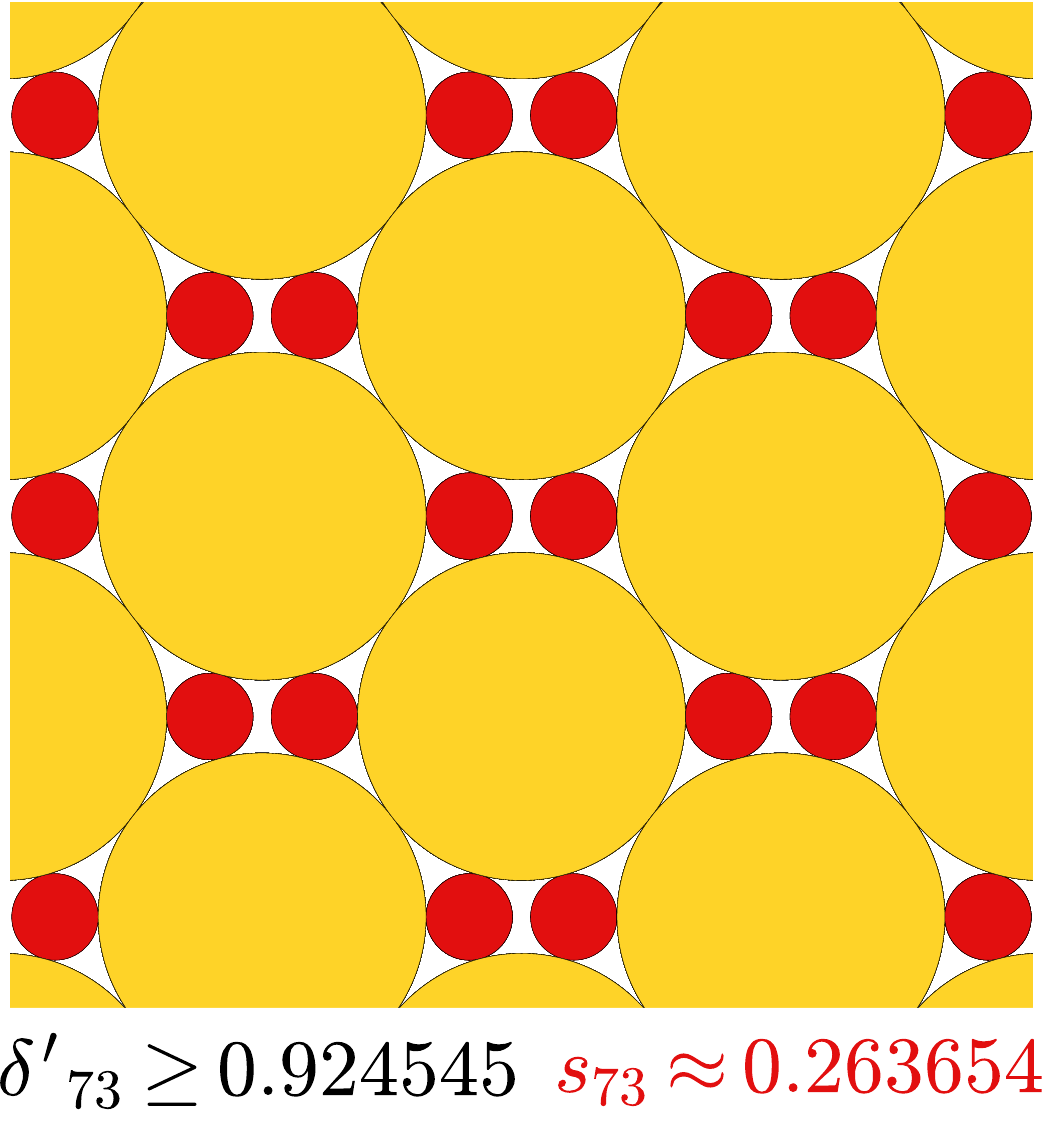}\hfill
  \includegraphics[width=.32\textwidth]{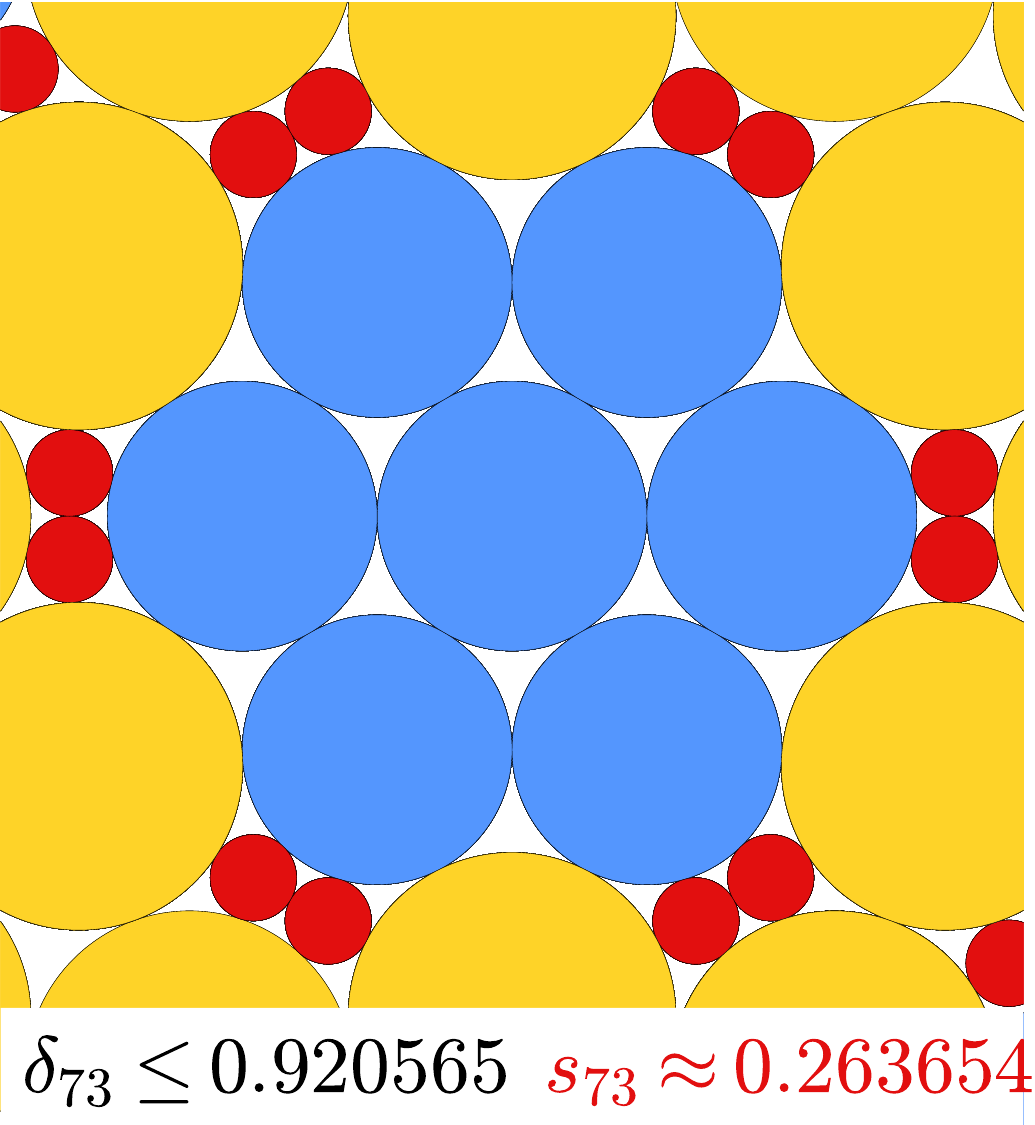}
  \caption{Left: a triangulated binary packing of case $b_7$. Middle:
    a deformation where the small discs are replaced with the small
    discs of case 73. Right: a triangulated periodic packing of case
    73, its fundamental domain and description are given
    in~\cite{164}.}
  \label{fig:73}
\end{figure}\vskip1ex

Let us explain our method on an example.  Recall that the pairs of
discs allowing binary triangulated packings are denoted by
$b_1,\dots,b_9$ while the triplets with ternary triangulated packings
are indexed by positive integers from $1$ to $164$.  Let us consider
case 73, its triangulated ternary packing is given in
Figure~\ref{fig:73}, on the right. Notice that the radius of the small
disc ($s_{73}\approx 0.263$) of case 73 is close to the radius of the
small disc ($r_{b_7}\approx 0.281$) of case $b_7$. Let us deform the
triangulated binary packing of $b_7$ (Figure~\ref{fig:73}, on the
left) replacing the small disc of $b_7$ by the small disc from
$73$. We choose a deformation which breaks as few contacts between
discs as possible: the one given in the center of Figure~\ref{fig:73}.
Observe that the only broken contact is between the two small discs:
they are not tangent anymore. The density of this new non-triangulated
packing $\delta'\approx 0.9245$ is higher than the density of the
triangulated packing $73$ $\delta_{73}\approx 0.9206$
(Figure~\ref{fig:73}, on the right).

This method is called flip-and-flow~\cite{flipflow}. The 40 counter
examples were found by computer search. First, for each case $b_i$, we
find the set of pairs of radii from the cases $1$-$164$ with radii
ratio ``close enough'' (we choose the distance heuristically) to the
ratio of the discs of $b_i$. Then we deform the triangulated packing
of $b_i$ to obtain packings with the found disc ratios. Our way to
deform packings was chosen in order to minimize the number of broken
contacts between discs since intuitively it is the best way to keep
the density high. Finally, the densities of 40 packings obtained by
our method were higher than the densities of the respective ternary
triangulated packings which leaves us with the counter examples given
in Appendix~\ref{app:counter}.

Our method is not universal: there might be other deformations for
certain cases to obtain even higher density and even more counter
examples. Besides that, there might be other cases with ternary
counter examples (notably, among the cases discussed in
Sections~\ref{sec:empty}).
\
\section{Other cases}
\label{sec:others}

\subsection{2 coronas}
\label{sec:twocoronas}

Among the necessary conditions on vertex potentials in tight triangles
given in Section~\ref{sec:vertex}, we saw that the sum of potentials
in the corona around any vertex of triangulated packing $T^*$ must be
equal to zero. In all the proved cases, each disc has only one
possible corona in $T^*$. It is not always the case, more precisely,
among the cases where $T^*$ is saturated, and for which we did not
find counter examples, there are 22 cases where one of the discs
appears with at least two different coronas in $T^*$: 16, 17, 18, 36,
49, 52, 57, 58, 65, 78, 84, 90, 106, 114, 120, 148, 153, 155, 156,
157, 158, 160. Each of these cases features a supplementary corona
consisting of 6 discs of the same size as the central one. We thus
have to add a supplementary condition $6V_{xxx}=0$, where $x$ is the
radius of the disc with two coronas. This however contradicts the
condition $3V_{xxx}=E_{xxx}$ in all of these cases. Our density
redistribution would need to be less local to solve this problem. In
the context of binary triangulated packings, such a case ($b_5$, see
Figure~\ref{fig:9_triangulated}) is treated in detail in Section~5.3
of~\cite{BF}.

\subsection{Empty polyhedra}
\label{sec:empty}
In Section~\ref{sec:poly}, we construct a polyhedron in $\RR^9$ aiming
to contain all valid values of tight vertex potentials and
$m_1,m_r,m_s$. In this Section, we talk about the 52 cases where the
polyhedron obtained by our computations is empty: 21, 22, 23, 26, 27,
34, 35, 46, 48, 50, 59, 61, 67, 68, 69, 71, 72, 74, 81, 82, 83, 85,
86, 87, 88, 89, 91, 94, 96, 101, 102, 103, 105, 107, 109, 112, 113,
121, 122, 123, 124, 125, 127, 128, 130, 134, 140, 143, 145, 147, 149,
150.

The polyhedron formed by the inequalities from
Section~\ref{sec:vertex}) and inequality~(\ref{ineq:eps}) for
$\epsilon=0$, represents the values satisfying~(\ref{eq:ver})
featuring a non-negative valid $\epsilon$. These constraints are
necessary for our proof to be correct. If this polyhedron is empty
there are no valid values of tight potentials and $m_1,m_r,m_s$ and
thus our strategy of proof is not applicable.

Nevertheless, our computations are limited by computer memory which
can represent only certain values. Normally, we avoid this problem by
using interval arithmetic (Section~\ref{sec:rif}). However, we can
not apply this solution with polyhedra. First, as mentioned in
Section~\ref{sec:poly}, in SageMath, the polyhedra module does not
support the interval field as a base ring. Implementing another way to
represent ``interval polyhedra'' would be unreasonable due to memory
and time constraints of calculations: the polyhedra are constructed
from thousands of inequalities, and performing computations in
interval field significantly increases time and memory costs. Instead,
we use the ring of rationals to store the inequalities
coefficients. Therefore, the polyhedron we work with is an
approximation of the actual polyhedron and may not contain all the
valid sets of values.

Yet, we believe that the polyhedra in question are probably actually
empty in these cases, so the precision issues are not the principal
obstacle.  All in all, some of the cases from this section might
actually maximize the density but we would need an essentially
different approach to be able to prove it. Looking forward, further
attempts to treat these cases would likely need to use a less local
density distribution.

\subsection{The 4 mysterious cases}
\label{sec:last}

\begin{figure}[htp]
  \center
  \includegraphics[width=.93\textwidth]{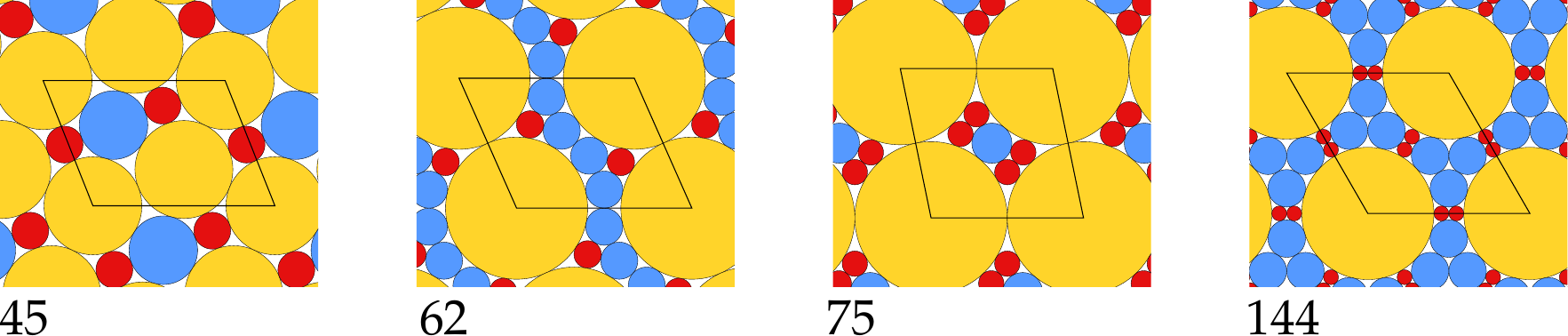}
  \caption{Triangulated ternary packings of the four \emph{mysterious}
    cases.}
  \label{fig:myst}
\end{figure}\vskip1ex

In the four remaining cases (45, 62, 75, and 144) the polyhedron from
Section~\ref{sec:poly} is not empty, like for the cases from the
previous section. Nevertheless, we could not find a point in it to
guarantee the local inequality~(\ref{ineq:tri}) in all triangles: the
problematic triangles are always those close to one of the tight
ones. Minimizing $m_q$ and the tight potentials is an obvious strategy
to minimize the potentials and eventually satisfy~(\ref{ineq:tri}) but
the capping constants $Z_q$ also dramatically affect potentials.

Trying to find appropriate values of $V_{xyz}, m_q$ and $z_q$, we
represented them with all the constraints coming from (\ref{ineq:U})
as a linear optimization problem.  This allowed us to encode
problematic triangles violating (\ref{ineq:tri}) as constraints and
add them to the system, one by one, each time one appears during local
verification (Section~\ref{sec:alltriangles}), in hope to finally
``converge'' to a solution which would satisfy (\ref{ineq:tri}) on all
triangles. However, this method failed: no solutions were found.

The fact that we could not choose a set of appropriate constants in
these cases does not prove that they do not exist (due to the
approximation issues already discussed in the previous section as well
as the new ones coming from encoding our constraints into a rational
linear problem). We, however, believe that these cases, as well as
those from the previous section, just can not be treated by our proof
methods. They probably require a less local emptiness redistribution
then the one we use. 

\bibliographystyle{alpha} 
\bibliography{3discs}

\appendix
\section{All counter examples}
In this section, we give all the found counter examples grouped by a
type of binary triangulated packing we used to construct it (the list
of binary triangulated packings is given in
Fig.~\ref{fig:9_triangulated}). Each counter example is presented as
in Fig.~\ref{fig:73}: to the left, we give the deformed binary packing
using a pair of discs whose radii ratio is close to the one of the
binary triangulated packing; to the right, the triangulated ternary
packings. \vspace{2ex}
\label{app:counter}


\newcommand{\binexamples}[2]{%
  \footnotesize%
  \begin{multicols}{2}%
    \begin{center}
    \foreach \x in #2 {%
      \noindent\stackunder[1pt]{%
        \includegraphics[width=.225\textwidth]{pictures/counter_ex_paper/\x_counter}
      }{\x \, counter example}
      \stackunder[1pt]{%
        \includegraphics[width=.225\textwidth]{pictures/counter_ex_paper/\x_ter}
      }{\x \, triangulated}\vspace{1ex}\hfill%
    }
    \end{center}
  \end{multicols}\vskip1cm
}

\subsection{Counter examples derived from b1}
\binexamples{b1}{{51,110}}

\subsection{Counter examples derived from b3}
\binexamples{b3}{{117,119}}

\subsection{Counter examples derived from b4}
\binexamples{b4}{{111}}

\subsection{Counter examples derived from b5}
\binexamples{b5}{{47,151}}

\subsection{Counter examples derived from b6}
\binexamples{b6}{{19, 63}}

\subsection{Counter examples derived from b7}
\binexamples{b7}{{60, 64, 70, 73, 80, 95, 98, 99, 104, 133, 137, 139, 142, 152, 154, 159, 161, 163}}

\subsection{Counter examples derived from b8}
\binexamples{b8}{{92, 97, 100, 126, 132, 136, 138, 162, 164}}

\subsection{Counter examples derived from b9}
\binexamples{b9}{{20, 25, 135, 141}}


\end{document}